# The Symmetry of the Boron Buckyball and a Related Boron Nanotube

N. Gonzalez Szwacki[1] and C. J. Tymczak[2*]

[1]*Institute of Theoretical Physics, Faculty of Physics, University of Warsaw, ul. Hoża 69, 00-681 Warsaw, Poland*

[2]*Department of Physics, Texas Southern University, Houston, Texas 77004, USA*

**Abstract**

We investigate the symmetry of the boron buckyball and a related boron nanotube. Using large-scale *ab-initio* calculations up to second-order Møller–Plesset perturbation theory, we have determined unambiguously the equilibrium geometry/symmetry of two structurally related boron clusters: the $B_{80}$ fullerene and the finite-length (5,0) boron nanotube. The $B_{80}$ cluster was found to have the same symmetry, $I_h$, as the $C_{60}$ molecule since its 20 additional boron atoms are located exactly at the centers of the 20 hexagons. Additionally, we also show that the (5,0) boron nanotube does not suffer from atomic buckling and its symmetry is $D_{5d}$ instead of $C_{5v}$ as has been described by previous calculations. Therefore, we predict that all the boron nanotubes rolled from the $\alpha$-sheet will be free from structural distortions, which has a significant impact on their electronic properties.

*E-mail: tymczakcj@tsu.edu

## 1. Introduction

Only recently the most energetically stable boron sheet, the so called $\alpha$-sheet [1], has been theoretically described. This sheet is closely related to the very stable boron fullerene, $B_{80}$, which is predicted to be the boron analog of the famous $C_{60}$ fullerene [2, 3]. The $\alpha$-sheet is also a precursor of boron nanotubes [4] whose theoretical study is very important in the light of the recent experimental verification [5]. The boron nanotubes had been investigated theoretically (see Refs. [6-8] and references therein) previous to the first boron tubular forms being synthesized. Additionally, small planar and quasi-planar boron clusters have also been extensively studied both experimentally and theoretically [9]. Together, these efforts have made possible the deeper understanding of the most likely stable structure of all-boron nanotubes, fullerenes and sheets, but more work still needs to be done.

The structural analogy between the $B_{80}$ and the boron nanotubes has been demonstrated in Refs. [10, 11]. Furthermore, Zope *et al.* have shown the link between $B_{80}$ and the $\alpha$-sheet [12]. Despite all the success of the theoretical description of the $B_{80}$ cluster, whose structure "inspired" many other investigations, the description of its symmetry is still controversial. The $B_{80}$ cluster was originally predicted to have the full icosahedral symmetry [2]. These calculations were done using the DFT-GGA approach. In a later publication, Gopakumar *et al.* have shown that the symmetry of the boron structure is not $I_h$, but instead, the cluster slightly distorts into the $T_h$ symmetry [13] where two such structural distortions, called **A** and **B**, have been identified at both Hartree-Fock (HF) and hybrid B3LYP levels of theory. The symmetry of $B_{80}$ was also addressed in several other later papers [14-16]. For instance, Sadrzadeh *et al.* demonstrated that in fact there is not one but three isomers, of $C_1$, $T_h$, and $I_h$ symmetries, which are close in energy and have almost identical structures [15].

The ambiguity in the description of the symmetry of the $B_{80}$ fullerene using pure DFT or hybrid approaches motivated us to investigate the structure of this cluster using the *ab-initio* second-order Møller–Plesset perturbation method (MP2). This method, although computationally expensive, is characterized by a much more accurate description of electron correlation effects than the DFT or hybrid HF/DFT methods can achieve. Additionally, we have extended our investigation to the finite-length boron (5,0) nanotube using the MP2 approach.

## 2. Computational details

The calculations have been carried out using both symmetry restricted and unrestricted methodologies. The computations with restricted symmetry have been done using the NWChem code suite [17]. We have done pure DFT (BLYP), hybrid DFT (B3LYP), HF, and MP2 calculations. The vibrational analysis and IR spectrum where obtained using tight convergence criteria. The symmetry unrestricted calculations have been done using the FreeON code suite [18]. The PBE, PBE0, and X3LYP functionals [19] have been used for these calculations. Additionally, the FreeON code suite has been used for nudged elastic band calculation (using the B3LYP functional) of the minimum energy paths (MEP) between two $B_{80}$ isomers of $T_h$ symmetry.

## 3. Results and discussion

### 3.1 Boron buckyball

Several tests for the structure and symmetry of $B_{80}$ have been performed at the MP2/STO-3G level of theory. First, we did several computations at the B3LYP/STO-3G level of theory. At this level, the total energy difference between clusters confined to $I_h$ and $T_h$ (isomer **A** from Ref. [13]) symmetries is $\Delta E$= 36.41 kcal/mol (see Table I), with the structure with $T_h$ symmetry being energetically more favorable. The 20 atoms that are located above or below the hexagonal rings of the $B_{60}$ frame are divided in two groups of 8 and 12 atoms. The 8 atoms are inside the frame with a dihedral angle 16.9° and the 12 atoms are outside the frame with a dihedral angle 7.2°. The optimized $T_h$-$B_{80}$ cluster is shown in Fig. 1(a) (left). In this figure both groups of hexagonal pyramid units are highlighted. The vibrational

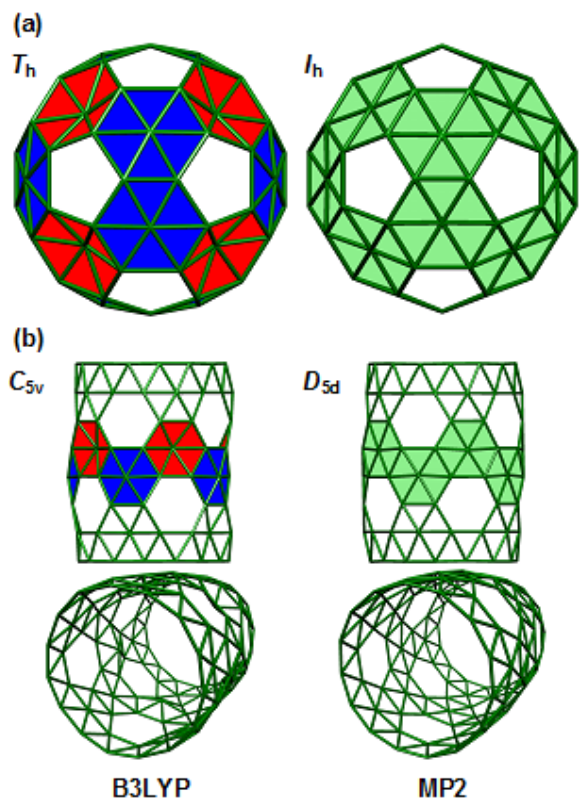

**FIG. 1 (a):** The $T_h$ and $I_h$ symmetries of the boron fullerenes, and **(b)** the $C_{5v}$ and $D_{5d}$ symmetries of the (5,0) boron nanotubes. The structures at the right (left) are optimized using the MP2/STO-3G (B3LYP/STO-3G) level of theory. The central atoms of the highlighted in red and blue units are closer or father away, respectively, from the center (or axis) of the cluster.

frequency analysis of the $I_h$ structure gives rise to 7 imaginary frequencies whose values range from $-287i$ to $-264i$ cm$^{-1}$. This picture is qualitatively, and in some cases quantitatively, similar to the previously reported calculations that had been done using the B3LYP/6-31G(d) level of theory [13].

The next step was to optimize the two clusters obtained at the B3LYP/STO-3G level of theory using the same bases set and the *ab-initio* MP2 approach. The optimization of the $T_h$ structure gave rise to a cluster of exactly the same inter-atomic bonds, symmetry, and total energy as the optimized $I_h$ structure (this is not surprising since $T_h$ is a subgroup of $I_h$). The optimized $I_h$-B$_{80}$ cage is shown in Fig. 1(a) (right). To validate this result, we have randomized the positions of the atoms of the $I_h$ cluster making small atomic displacements and after re-optimization the inter-atomic distances and angles where within 0.001 Å and 0.1°, respectively, of the $I_h$ cluster.

In Fig. 2 we have shown the MEP between two B$_{80}$ isomers of $T_h$ symmetry described in Ref. [13]. The cluster images (structures) where first calculated at the B3LYP/STO-3G level of theory and then for these structures single energy calculations using HF, BLYP, and MP2 methods where done. Interestingly enough, although the B3LYP images are not the true images for the other methods, still we obtained correct energy barriers for the transition from isomer **A**, through $I_h$ symmetry, to isomer **B**. We found the energy barriers to be ~106 and ~25 kcal/mol for the HF and BLYP methods, respectively (see

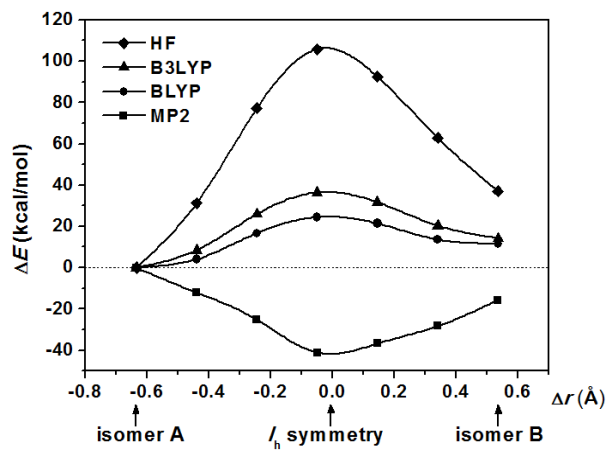

**FIG. 2:** The transition states for the **A** to **B** isomers for HF, BLYP, B3LYP, and MP2 theory levels using a STO-3G basis set. $\Delta r$ is the difference between the radial distances of the boron atoms belonging to the two groups of 8 and 12 atoms located near the centers of the hexagonal rings of the B$_{60}$ frame.

Table I). As is also shown in Fig. 2 the HF, BLYP, and B3LYP theories predict that the ground state structure is isomer **A**, whereas, in contrast, the MP2 theory clearly predicts that the ground state structure is of $I_h$ symmetry with an energy ~26 kcal/mol bellow the energy for isomer **B**.

It is noteworthy to mention that the transition states energy barriers shown in Fig. 2 decrease as basis set completeness is approached for the HF and DFT theories. Considering Table I, we see that the $\Delta E$ values (energy barriers) for the BLYP or B3LYP functionals decrease with increasing basis set completeness, e.g. the aug-cc-pVDZ basis set energy barriers are very small which implies very flat energy surfaces along the MEPs. This is more likely the reason for the coexistence of several low-lying isomers that are very close in energy as reported in the literature [20]. We believe that the ambiguity in the description of the structure of B$_{80}$ is a consequence of correlation effects that cannot be fully captured by DFT. It is well known that DFT does not always accurately describe correlations, especially Van der Waals interactions [19]. If we assume that part of the attractive interaction between the central boron atom and the six-member boron ring is of dipole-dipole character, then DFT will fail to capture these interactions effectively, and higher-level correlation theories have to be used in order to correctly predict the structure of B$_{80}$.

To ensure that our predictions are independent on the choice of the basis set, we have repeated the MP2 calculations but now with the 4-31G basis set. The starting point for those calculations were the structures obtained at the B3LYP/4-31G level of theory. Again the optimized $T_h$ and $I_h$ clusters have exactly the same structure and energy. Our last step, in the MP2 calculations, was to optimize the $I_h$ structure at the MP2/6-31G(d) and MP2/cc-pVDZ levels of theory. At those levels, the B$_{80}$ cluster was found to have the same topology as the C$_{60}$ molecule since the 20 additional boron atoms are located almost exactly at the centers of the 20 hexagons of the B$_{60}$ frame.

We summarize our results in Table I where we show the equilibrium inter-atomic distances of $I_h$-B$_{80}$ for HF, BLYP, B3LYP and MP2 levels of theory at increasing completeness of the basis set. We also list the energy difference, $\Delta E$, between $B_{80}$ clusters of $I_h$ to $T_h$ symmetries. From columns 3-5 of Table I, we can see that HF, BLYP, and B3LYP methods clearly underestimate the hexagon-hexagon ($d_{hh}$) and overestimate the hexagon-pentagon ($d_{hp}$) B-B distances of the $B_{60}$ frame respect to the results obtained using the MP2 method. However, all the approaches tend to give similar results for the distance, $d_c$, from an atom of the hexagonal ring to the central boron atom. It is worth mentioning that the MP2/STO-3G and MP2/6-31G(d) theories predict that all the B-B inter-atomic distances are close to 1.71 Å, which is the bond length of the flat boron triangle sheet [1, 10]. We can also see from Table I that even for a very large basis set, aug-cc-pVDZ, the theories, BLYP and B3LYP, predict that the $T_h$ symmetry corresponds to the equilibrium geometry of the cages (the HOMO-LUMO gaps are 0.99 and 1.93 eV at the BLYP/aug-cc-pVDZ and B3LYP/aug-cc-pVDZ levels of theory, respectively). However, it should be pointed out that such small energy difference, between the $I_h$ and $T_h$ structures, is at the limit of the accuracy of the DFT and hybrid approaches [21]. And finally we again mention that the MP2 theory, for all the basis sets we have used, predict that the $I_h$ symmetry is the ground state symmetry of the boron buckyball. In Fig. 3, we show the IR spectrum of the $B_{80}$ cage calculated at the MP2/4-31G level of theory. All the obtained vibrational frequencies are, contrary to the B3LYP/4-31G case, positive which is an additional indication of the stability of the $I_h$-$B_{80}$ cluster. The infrared peaks at frequencies 372, 425, and 874 cm$^{-1}$ correspond to weak IR modes that are localized on the $B_{60}$ frame and are equivalents of the $T_{1u}(1)$, $T_{1u}(2)$ and $T_{1u}(3)$ modes, respectively, present in the infrared spectrum of the $C_{60}$ fullerene [22]. The strong modes at frequencies 329, 741 and 968 cm$^{-1}$ involve all the 80 atoms of the boron cage, although the first frequency (329 cm$^{-1}$) is mainly localized on 20 atoms that are located in the centers of the hexagons of the $B_{60}$ frame.

TABLE I: Equilibrium inter-atomic distances of $I_h$-$B_{80}$ at various levels of theory. $d_{hh}$ and $d_{hp}$ refer to hexagon-hexagon and hexagon-pentagon bond length, respectively. $d_c$ refers to the distance from an atom of the hexagonal ring to the central boron atom. The six-member ring and the central atom define a dihedral angle that is also provided. A negative (positive) angle means that the central atom is shifted towards (away from) the center of the cage. $\Delta E$ is the total energy difference between clusters confined to $I_h$ and $T_h$ (isomer **A**) symmetries.

|  | $\Delta E$ (kcal/mol) | NN interatomic distances (Å) | | | dihedral angle |
|---|---|---|---|---|---|
|  |  | $d_{hh}$ | $d_{hp}$ | $d_c$ |  |
| HF | | | | | |
| STO-3G | 106.42 | 1.607 | 1.752 | 1.710 | -11.3° |
| 4-31G | 63.12 | 1.652 | 1.762 | 1.716 | -6.5° |
| cc-pVDZ | 14.87 | 1.656 | 1.774 | 1.731 | -8.7° |
| BLYP | | | | | |
| STO-3G | 24.68 | 1.675 | 1.748 | 1.719 | -5.9° |
| 4-31G | 7.89 | 1.684 | 1.739 | 1.712 | -1.1° |
| cc-pVDZ | 0.13 | 1.691 | 1.751 | 1.724 | -3.6° |
| aug-cc-pVDZ | 0.20 | 1.688 | 1.749 | 1.721 | -3.4° |
| B3LYP | | | | | |
| STO-3G | 36.41 | 1.654 | 1.741 | 1.707 | -6.8° |
| 4-31G | 13.46 | 1.669 | 1.732 | 1.701 | -1.4° |
| cc-pVDZ | 0.15 | 1.677 | 1.746 | 1.715 | -4.3° |
| aug-cc-pVDZ | 0.11 | 1.672 | 1.745 | 1.712 | -4.3° |
| MP2 | | | | | |
| STO-3G | 0.0 | 1.709 | 1.713 | 1.712 | -1.7° |
| 4-31G | 0.0 | 1.735 | 1.713 | 1.725 | +2.1° |
| 6-31G(d) | * | 1.711 | 1.706 | 1.708 | 0.0° |
| cc-pVDZ | * | 1.745 | 1.704 | 1.725 | +1.5° |

(*) only the $I_h$-$B_{80}$ cluster has been optimized

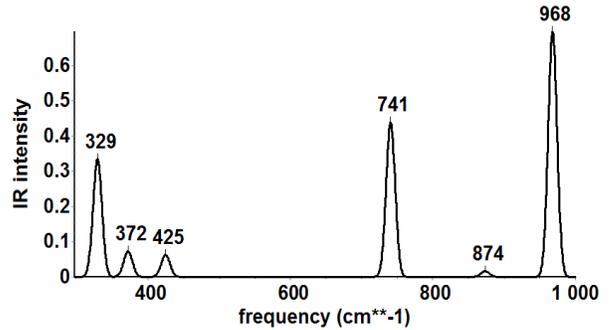

**FIG. 3:** The IR spectrum of the $I_h$-$B_{80}$ fullerene. The vibrational frequencies have been obtained at the MP2/4-31G level of theory.

For completeness of our investigation, we have also done computations unrestricting the symmetry of the structures. For those calculations, we have used several functionals, 6-31G(d) and 6-311G(d) basis sets, and tight convergence criteria. The results are essentially the same as reported in Ref. [15]. Depending on the symmetry of the input structures, after structural optimization we have obtained clusters with symmetries close to $T_h$ or $I_h$. Although the energy differences between those structures were very small (within 2.7 kcal/mol), the $T_h$ cages were always the lowest in energy.

### 3.2 Boron nanotubes

First principle calculations predict that boron nanotubes are metallic or semiconducting depending on their radii [4, 20]. The band gap depends also on the chirality of the tube [4]. The presence of the band gap in boron nanotubes with diameters smaller than 17 Å is associated with the buckling of certain number of boron atoms which occurs when the $\alpha$-sheet is rolled to obtain the nanotubes [4]. Structures with smaller radii tend to have larger structural distortions (buckling of atoms) and as a consequence larger band gaps [4, 20]. It was also shown that without this buckling the nanotubes would be metallic [11, 20]. The close similarity between the structural distortions that suffer boron fullerenes and nanotubes motivated us to investigate the structure of a finite-length (5,0) boron nanotube using the MP2 approach. This nanotube was reported to have one of the largest band gaps (~0.6 eV) among all studied nanotubes [4, 20]. It is also closely related to the $B_{80}$ cluster [10, 11]. It should be noted that in the literature we can find three different indexing schemes for boron nanotubes and, as a result, there are three different notations for the same nanotubes: (5,5) in Ref. [20], (5,0) in Refs. [4, 11], and (15,0) in Ref. [10]. In this work we use the $(n,0)$ notation for zigzag nanotubes which is consistent with the indexing scheme adopted in Ref. [6].

For our investigation, we chose a fragment of the (5,0) boron nanotube long enough, ~10 Å, to reproduce the structural properties of an infinite nanotube, but still computationally feasible. This finite-length nanotube, of

110 atoms, is shown in Fig. 1(b). To minimize the effects of the edges on the calculations the tubular fragment has boron double rings at the edges. The cluster has been first optimized at the B3LYP/STO-3G level of theory. The resulting structure, of $C_{5v}$ symmetry, is shown in the left part of Fig. 1(b). In this figure, we have highlighted a central fragment of the cluster which has a diameter of 8.47 Å (B3LYP/431G: 8.44 Å; B3LYP/cc-pVDZ: 8.42 Å), a close value to that reported for an infinite nanotube [20]. We can also see from that figure that there are two groups of nonequivalent hexagonal pyramid like units, which are building blocks of the nanotube (this was also reported for an infinite (5,0) nanotube [4, 11, 20]). Next, we have farther optimized this structure at the MP2/STO-3G level of theory and found that the equilibrium structure is of $D_{5d}$ symmetry (see Fig. 1(b) (right)). We have done several tests to ensure that the predicted equilibrium structure is a true local minimum of energy. The increase in cluster symmetry is accompanied by a flattening of the hexagonal pyramid units that are now all equivalent. This result may have important consequences on the electronic properties of boron nanotubes. We predict that not only the (5,0) nanotube is metallic, but it is very likely that the rest of the boron nanotubes that were previously classified as semiconducting do not suffer from structural distortions (buckling of atoms) and as a consequence are, in fact, metallic. This is important for nanotechnology since it indicates that boron nanotubes would be superior candidates, as opposed to carbon nanotubes, for electrical interconnects in nano-electronics.

At the limit of a very rich basis set the B3LYP results tend to be qualitatively the same as those obtained using the MP2 method and the minimal basis set. Indeed, the energy difference between the structures of $D_{5d}$ and $C_{5v}$ symmetries is only 2.34 kcal/mol at the B3LYP/cc-pVDZ level of theory, whereas for the STO-3G and 4-31G basis sets the values are 40.02 and 10.33 kcal/mol, respectively. This result is expected since more complete basis sets favor better description of correlation effects. Finally, it should be mentioned that the HOMO-LUMO gap slightly decreases (B3LYP/cc-pVDZ: from 0.81 to 0.79 eV for clusters of $C_{5v}$ and $D_{5d}$ symmetry, respectively) with increasing symmetry of the tubular cluster.

## 4. Conclusions

In conclusion, we have done extensive calculations at various levels of theory to determine the equilibrium geometry of the $B_{80}$ cage and a related boron nanotube. We have determined that the equilibrium geometry of $B_{80}$ is $I_h$, the same as for $C_{60}$, and of the boron nanotube is $D_{5d}$. From these results, we have asserted that a high level description of the correlation effects is essential for the correct description of the structure of these and other hollow boron nanostructures.

## Acknowledgment

We would like to thank the Robert A. Welch Foundation (Grant J-1675) for their support of this project. The authors would also like to acknowledge the High Performance Computing Center (URL: http://hpcc.tsu.edu/) at Texas Southern University for providing resources that have contributed to the research results reported within this paper.